\documentstyle[aps,prl,preprint,floats,epsfig]{revtex}

\begin{document}
\draft

\preprint{\tighten\vbox{\hbox{\hfil CLNS 97/1517}
                        \hbox{\hfil CLEO 97-24}
}}

\title{ Measurement  of \boldmath{$Br(D^0\to K^-\pi^+)$} using 
partial reconstruction of \boldmath{$\overline B \to D^{*+}X\ell^-\bar\nu$} }  

\author{CLEO Collaboration}
\date{\today}

\maketitle
\tighten

\begin{abstract} 
We present a measurement of the absolute branching fraction for 
$D^0\to K^-\pi^+$ using  the reconstruction of the decay chain  
$\overline B \to D^{*+} X \ell^-\bar\nu$, $D^{*+} \to D^0 \pi^+$ 
where only the lepton  and the low-momentum pion from the 
$D^{*+}$ are  detected.
With data collected by the CLEO II detector at the Cornell Electron
Storage Ring, we have determined  $Br(D^0\to K^- \pi^+) = [3.81\pm 0.15(stat.)\pm 0.16(syst.)]\%$. 
\end{abstract}

 
\newpage
{
\renewcommand{\thefootnote}{\fnsymbol{footnote}}

\begin{center}
M.~Artuso,$^{1}$ F.~Azfar,$^{1}$ A.~Efimov,$^{1}$
M.~Goldberg,$^{1}$ D.~He,$^{1}$ S.~Kopp,$^{1}$
G.~C.~Moneti,$^{1}$ R.~Mountain,$^{1}$ S.~Schuh,$^{1}$
T.~Skwarnicki,$^{1}$
% ...
\end{center}
 
\small
\begin{center}
$^{1}${Syracuse University, Syracuse, New York 13244}\\
% ...
$^{13}${University of Illinois, Urbana-Champaign, Illinois 61801}\\
$^{14}${Carleton University, Ottawa, Ontario, Canada K1S 5B6 \\
and the Institute of Particle Physics, Canada}\\
% ...
$^{25}${Southern Methodist University, Dallas, Texas 75275}
\end{center}

\setcounter{footnote}{0}
}
\newpage

As most of the published branching fractions of $D^0$, $D^+$ and $D_s^+$ mesons
are normalized to the $D^0 \to K^- \pi^+$ \cite{charge-conjugate} decay mode, then 
the value of  $Br(D^0 \to K^- \pi^+)$  directly affects many topics  in heavy flavor physics. 
Some examples include  charm counting in $B$ meson decays where about 90\% of the total 
charm  yield is calibrated by $Br(D^0 \to K^- \pi^+)$ \cite{CLEO-BDX}, the determination 
of $Br(Z^0\to c\bar c)$, and the  investigation of  any  exclusive decay  
mode of the $B$ meson which contains $D^0$, $D^+$ or $D_s^+$  in the final state.

In order to measure the absolute branching fraction for $D^0 \to K^- \pi^+$ decay, 
one needs to find the number of $D^0$'s without reconstructing a particular $D^0$ 
decay mode.
In this Letter  we present a measurement of the absolute
$D^0 \to K^- \pi^+$ branching fraction, developing the method  first
used by the ARGUS Collaboration \cite{ARGUS-Kpi-partial}. 
The inclusive number of $D^0$'s is determined by  partial
reconstruction of the decay chain $\overline B^0 \to D^{*+} \ell^-
\bar\nu$, $D^{*+} \to D^0 \pi^+$, where only the lepton  and the
slow pion  from the $D^{*+}$, hereafter denoted as $\pi_s$, are detected. 
The systematic errors involved are largely different from those of other recent  
measurements \cite{ALEPH-Kpi,ARGUS-Kpi,CLEO-Vivek}, where slow pions  within jets were used to tag the 
decay  $D^{*+} \to D^0 \pi^+$.

We have used 3.1 $fb^{-1}$ of data collected on the $\Upsilon(4S)$ resonance by 
the CLEO II detector \cite{CLEOII}.
The data set corresponds to $3.3 \times 10^6$ $B \overline B$ events. 
In order to suppress non-$B \overline B$ (continuum) background we required the ratio of the 
Fox-Wolfram moments $H_2/H_0$ \cite{Fox-Wolfram} to be less than 0.4.
The remaining contribution from continuum events was estimated  using
1.6 $fb^{-1}$  of data collected just below the $B \overline B$ threshold. 
In the following this continuum subtraction is implicit.

We required lepton candidates to have a momentum  between 1.4 GeV/$c$ and 2.5 GeV/$c$
and to be in the barrel region of the detector. 
Muon candidates were required to penetrate an iron absorber to a depth of at least
5 nuclear interaction lengths.
Electrons were identified through  a comparison of the energy deposited in the 
electromagnetic calorimeter with  the momentum measured in the drift chambers
and by specific ionization energy loss ({\tt dE/dx}) measurements.  
 We required that the $\pi_s$ candidate  have a momentum lower than  190 MeV/$c$,  
which is slightly below the upper kinematic limit for  pions  from  $D^{*+}$ in 
$\overline B \to D^{*+} \ell^- \bar\nu$ decays.
      
The partial  reconstruction of the decay $\overline B \to D^{*+} \ell^- \bar\nu$ 
 exploits the extremely low energy release in the decay 
$D^{*+} \to D^0 \pi_s^+$.
The pion  is almost at rest in the $D^{*+}$ frame, and its  velocity vector
in the lab frame  is approximately 
equal to that of the $ D^{*+}$.
Our main signal mode is $\overline B^0 \to D^{*+} \ell^- \bar\nu$, for which
the missing mass squared is calculated as
\begin{eqnarray}
 MM^2   =  (E_{B}-E_\ell-E_{D^{*+}})^2-|\vec P_B- \vec P_\ell - \vec P_{D^{*+}}|^2.
\label{eqn:MM} 
\end{eqnarray}
The energy of the $B$ meson is precisely the  beam energy. We do not know the direction of motion of the $B$,
but the $B$ momentum is sufficiently small  ( $\approx$ 300 MeV/$c$) compared to the  typical 
values of $|\vec P_\ell |$ and $|\vec P_{D^{*+}}|$ that we can  
set  $\vec P_B=0$. We approximated the direction of motion of the 
$D^{*+}$ by the direction of motion of the $\pi_s$.
If the $\pi_s$ were exactly at rest in the  $D^{*+}$ frame, the $D^{*+}$ energy would be given by  
$E^{lab}_{D^{*+}}=(E^{lab}_{\pi}/E_{\pi}^{c.m.}) \cdot M_{D^{*+}}$.
In order to  correct for the non-zero momentum 
of the  $\pi_s$  in the  $D^{*+}$ frame,  we used
a parameterization obtained from Monte Carlo to estimate $E_{D^{*+}}$ as a function 
the   $\pi_s$ momentum \cite{improved-partial}.

\begin{figure}[p]
\centering
\epsfig{figure=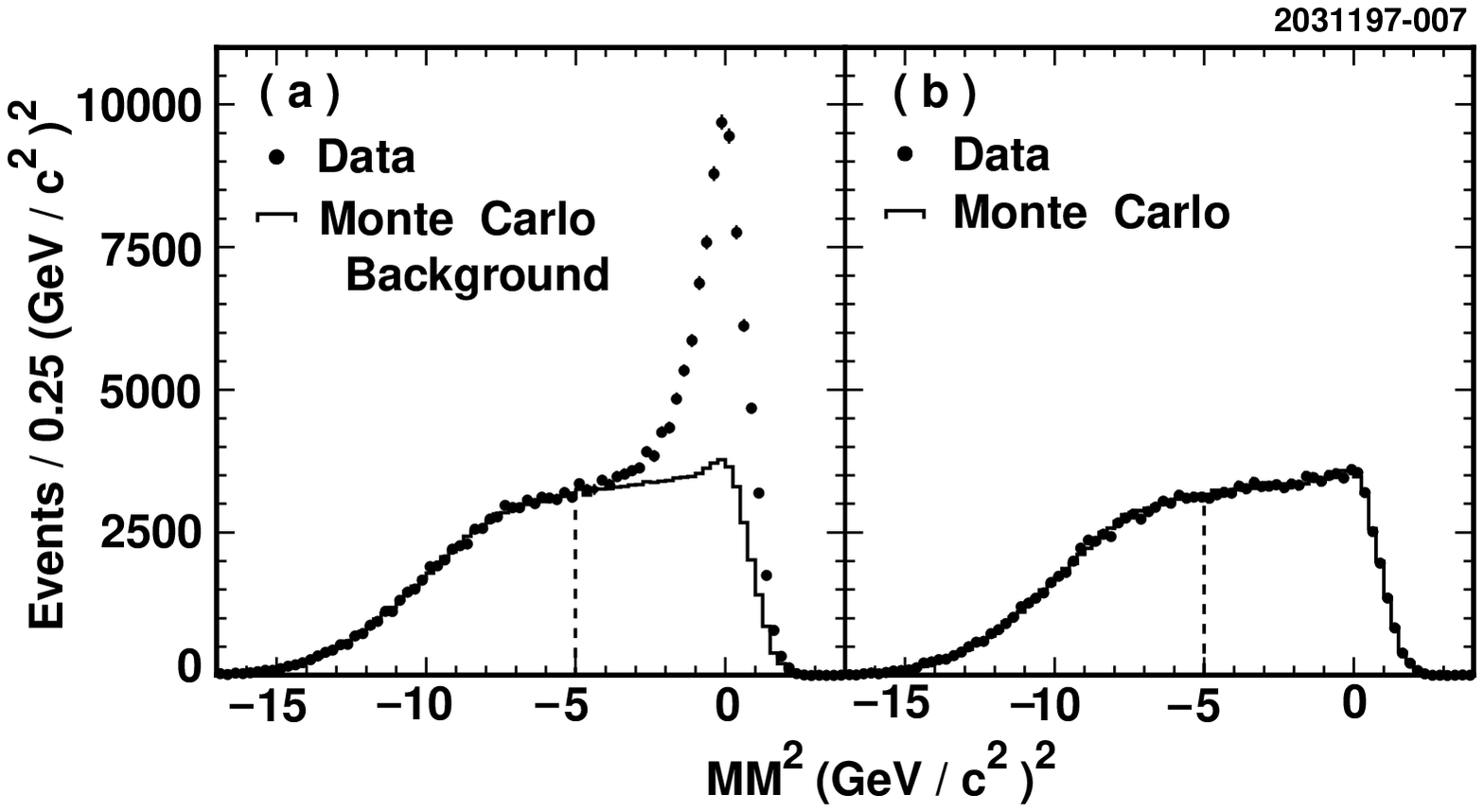}
\caption {The missing mass squared ($MM^2$)  distribution for the right-sign {\it (a)} and wrong-sign {\it (b)}
 $\ell \: \pi_s$ pairs. The Monte Carlo background shape  has been normalized to  
the data distribution  in the sideband region indicated  by the dashed line 
($MM^2<-5$ GeV$^2/c^4$). }
\label{fig:paper_mnu}
\end{figure}

The resulting $MM^2$ distribution
is shown in Figure~\ref{fig:paper_mnu}(a).
The events with the lepton and slow pion coming from  $\overline B^0 \to D^{*+} \ell^- \bar\nu$, 
$D^{*+} \to D^0 \pi^+_s$ produce a prominent peak  at $MM^2 \approx 0$.
However, the decays $\overline B\to D^{*+} X \ell^-\bar\nu$,
$D^{*+} \to D^0 \pi_s^+$  also contribute to 
this peak. We have considered these decay modes to be  signal because they 
produce true $D^{*+} \to D^0  \pi_s^+$.
 More specifically, we allowed the $D^{*+}$ to come from 
$\overline B \to  D^{*+} n\pi  \ell^-\bar\nu$ decays, where  $D^{*+} n\pi$ may or may not form a resonance. We  also allowed the lepton to come from $\tau$  in the decays $\overline B \to D^{*+} \tau^-\bar\nu$
or from ${\overline D}$ in the decays $\overline B \to D^{*+} {\overline D} X$, where $\overline D$  represents 
$\overline D^{0}$, $D^{-}$ or $D_{s}^{-}$. Our analysis is  therefore   not dependent on the branching fractions 
assumed in the Monte Carlo  for  the poorly measured $\overline B \to  D^{*+} n\pi  \ell^-\bar\nu$  
and  $\overline B \to D^{*+} \overline D X$  decays, because these decays were considered to be  signal.

A  Monte-Carlo  simulation of the $B \overline B$ events was  used  to  determine the 
background shape. We  normalized the  background shape to the data distribution in the sideband  region ($MM^2<-5$ GeV$^2/c^4$).
After the background  subtraction, the number 
of events in the signal region (defined as $MM^2>-2$ GeV$^2/c^4$ ) was found to be $N^{incl}=44,504\pm360\:(stat.)$. In this way we have  extracted the  number of  $\overline B \to D^{*+} X \ell^- \bar\nu$ events in which  $D^{*+} \to D^0 \pi_s^+$. 

\begin{figure}[p]
\centering
\epsfig{figure=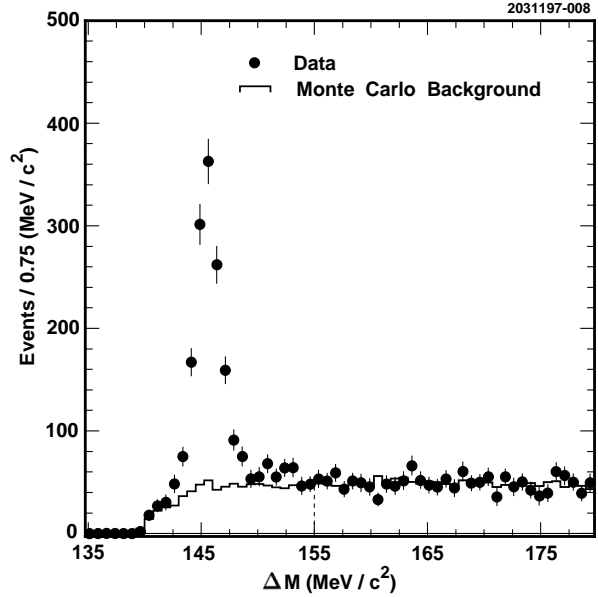,width=8cm,height=8cm}

\caption { $\Delta M \equiv M( K^- \pi^+\pi_s^+)-M( K^- \pi^+)$ distribution for data 
with the Monte Carlo 
background shape normalized  to the  data  distribution in the sideband region. 
The lower limit for the sideband region is  indicated by the dashed line. }
\label{fig:paper_mdiff}
\end{figure}

We  have thus obtained a sample of 
$D^{*+} \to D^0 \pi^+$ decays  without reconstructing  a particular $D^0$ decay mode. 
Next we need to determine how many $D^0$'s from these $D^{*+} \to D^0 \pi^+$ 
events decay to   $K^- \pi^+$.
For every $\ell^-  \pi_s^+$ pair for which the value of  $MM^2$ was within the signal region  
($MM^2>-2$ GeV$^2/c^4$)   we  searched 
for a  $K^- \pi^+$ pair, assigning the kaon mass to the track of the  opposite charge 
with respect to $\pi_s$, and   requiring $|M(K^- \pi^+)-M(D^0)|<35$ MeV$/c^2$, which corresponds to  a $ 3.5\,\sigma$  cut. The $ K^- \pi^+$ pair was  combined  with the $\pi_s^+$ and  the  
mass difference    $\Delta M \equiv M( K^- \pi^+\pi_s^+)-M( K^- \pi^+)$ was formed. 
The resulting $\Delta M$ distribution is shown in Figure~\ref{fig:paper_mdiff}. The 
prominent peak at $\Delta M = M(D^{*+})-M(D^0)  \approx 145.4$ MeV$/c^2$  is produced 
by  $D^{*+} \to D^0 \pi^+$, $D^0\to K^-\pi^+$ decays.
We  normalized    the  background shape obtained from  the  Monte-Carlo simulation 
to the data  distribution in the sideband  
region (155 MeV$/c^2 < \Delta M<$180 MeV$/c^2$). 
True $D^{*+} \to D^0 \pi_s^+$, $D^0\to K^-\pi^+$ decays  where the  $D^{*+}$ does not come 
from a signal decay chain were considered to be background.
After the background subtraction we counted  the number 
of events in the signal region, defined as  141.50 MeV$/c^2< \Delta M<$149.75 MeV$/c^2$.
The number of  decays $D^{*+} \to D^0 \pi^+$     
with $D^0 \to K^- \pi^+$, denoted as $N^{excl}$, was found to be  $1165\pm 45 \:(stat.)$.

To extract $Br(D^0\to K^-\pi^+)$ we need to correct the ratio $N^{excl}/N^{incl}$ 
for the track  reconstruction  and acceptance efficiencies :
\begin{eqnarray}
Br(D^0\to K^-\pi^+)=\frac{N^{excl}}
{N^{incl}}\cdot
\frac{1}{\epsilon}.
\label{eqn:branching}
\end{eqnarray}
We obtained  $\epsilon$  using a GEANT-based  Monte-Carlo simulation \cite{GEANT} of the CLEO II detector.
To a good approximation the lepton and slow pion reconstruction efficiencies cancel  in the ratio when we calculate $\epsilon$. 
Therefore $\epsilon$ mainly includes reconstruction and selection efficiencies for 
 $K^-$ and $\pi^+$ tracks and   acceptance efficiencies for the $M(K \pi)$ and  
$\Delta M$ signal regions.
However, the cancellation of the lepton and slow pion reconstruction efficiencies is not exact  
because the average charged track multiplicity  for  $D^0$ decays is higher than that 
for $D^0\to K^-\pi^+$ mode and  it is more difficult to reconstruct a track in a higher multiplicity environment.
We found that this effect changes $\epsilon$ by $3.7\%$ of itself. In order to take this 
into account,  we calculated  $\epsilon$ by selecting  signal events from the Monte-Carlo simulation of  $B \overline B$ events, and 
comparing the value of $N^{excl}_{MC}/N^{incl}_{MC}$ to the branching ratio that was used in the Monte Carlo.  
Note that in this procedure  $N^{incl}_{MC}$ corresponds to the number
of $\overline B \to D^{*+}X\ell^-\bar\nu$, $D^{*+} \to D^0 \pi^+$ events where 
$D^0$'s were allowed to decay generically, not forced to  decay into  $K^-\pi^+$. 
We obtained  $\epsilon=[68.6 \pm 2.1(syst.)]\%$, and using this value of $\epsilon$ together with Eqn.~\ref{eqn:branching}, we  found 
\begin{displaymath}
Br(D^0\to K^- \pi^+) = [3.81\pm 0.15(stat.)\pm 0.16(syst.) ]\%.
\end{displaymath}

The total systematic error was obtained by summing in quadrature the errors given in  Table~I. 
We will now discuss the systematic uncertainties dividing the possible sources 
into three categories:  (i) determination of  $N^{incl}$ 
using  the  $MM^2$ distribution,
(ii) determination of  $N^{excl}$  using   the $\Delta M$ distribution, 
(iii) efficiency extraction from Monte Carlo.  

(i)  First, to  see how well the Monte Carlo can simulate 
the background shape for the  $MM^2$ distribution, we looked at the $MM^2$
distribution  for the wrong-sign (i.e. same sign)  $\ell \: \pi_s$ pairs 
(Figure~\ref{fig:paper_mnu}(b) ).
We normalized the Monte-Carlo shape  to  data distribution  in the sideband region ($MM^2<-5$ GeV$^2/c^4$), 
as we did for the right-sign $\ell \: \pi_s$ pairs,  and compared the Monte-Carlo prediction 
with data in the  signal region  ($MM^2 >-2$ GeV$^2/c^4$). We found excellent agreement 
within the statistical precision of  0.8\% of the signal region population. We include 
this   0.8\% as a part of the systematic error.
This result is encouraging, but different physics  can contribute to the distributions 
for wrong-sign  and right-sign background $\ell \: \pi_s$ pairs.
Using Monte Carlo, we performed a thorough study comparing the $MM^2$ distributions for
the various physical processes  producing   the wrong-sign or  
the right-sign background   $\ell \: \pi_s$ pairs.

We have found that the most dangerous source of background which peaks 
in the signal region of $MM^2$ distribution is the decay chain 
 $\overline B \to D X \ell^- \bar\nu$,  $D \to \hbox{(something heavy)}+\pi^+$, where the $\pi^+$
is moving slowly in the $D$ rest frame and  mimics the pion from  $D^{*+} \to D^0 \pi_s^+$ decay.
These decays do not contribute to the 
$\Delta M$ peak and thus can reduce the measured  $D^0\to K^-\pi^+$ branching fraction. 
To estimate the systematic error due to this background 
we identified  three such low $Q$-value decay modes  in our Monte Carlo: 
$D^{+} \to \overline{K}^{*0}(892) \omega \pi^+$, $D^{+} \to \overline{K}^{*-} \rho^+ \pi^+$, 
and $D^{+} \to \overline{K}^{*0} \rho^0 \pi^+$.
Monte Carlo  predicts that the events with the pion coming from one of these modes account for 
$0.7\%$ of the events under the  $MM^2$ peak  with respect to the number of events in the 
signal peak.
We have exploited   the difference  in the $MM^2$  distribution shapes for this background  
and the  signal and 
fit the whole $MM^2$ data distribution with three histograms obtained from  Monte Carlo:
signal, the contribution from the decay chain $\overline B \to D^{+}X\ell^-\bar\nu$ where $D^{+} \to \overline{K}^{*}  (\omega \: or \: \rho) \pi^+$, and  the rest of background. 
The fit  showed that  the contribution 
from these modes is consistent with the Monte-Carlo prediction. However 
we should keep in mind that the decay modes we are considering here are poorly  
measured  and that  there could be  other similar low $Q$-value decays that have  not yet been 
observed. In order  to be conservative, we  varied the  contribution 
from  $\overline B \to D^{+}X\ell^-\bar\nu$, $D^{+} \to \overline{K}^{*} (\omega \: or \: \rho) \pi^+$
 in the Monte-Carlo background shape by the fit error  and  obtained a $2.3\%$ variation in final result, 
which we took as the systematic error due to this background. This is the largest single source 
of systematic uncertainty in the analysis. 

Another  source of background which peaks 
in the signal region of the $MM^2$ distribution results when the slow pion from
a signal decay chain decays in flight to a  muon, and we identify this muon as the slow pion.
Monte Carlo predicts the magnitude of background from this source in the $MM^2$ peak region to be
$2.5\%$ of the signal.
Even though this is the largest source of background which peaks 
in the signal region it  does not significantly bias the $Br(D^0\to K^-\pi^+)$ measurement 
because this background produces smeared  peaks in the signal regions
of  both the $MM^2$ {\it and} the $\Delta M$ distributions. 
We varied the Monte-Carlo prediction for this  background by 30\% of itself
and obtained $0.3\%$ variation in final result, which we took as the systematic error. 

Another background which peaks 
in the $MM^2$ signal region    results
when we identify as a $\pi_s^+$ a positron  from  $\pi^0 \to \gamma e^+ e^-$ or 
$\gamma$ conversion in the decay chain $\overline B \to D^{*}X\ell^- \bar\nu$, 
$D^{*} \to D \pi^0$, $D \gamma$.
Monte Carlo predicts the magnitude of background from this source in the $MM^2$ peak region to be
$0.7\%$ of the signal.
We varied the Monte Carlo prediction for this  background by 30\% of itself
and obtained $0.4\%$ variation in final result, which we took as the systematic error.

 Combining the errors described  above in  (i) we estimated the systematic error due  to background subtraction in the the  $MM^2$ distribution to be 2.5\%.
We have also studied the possible systematic errors due to the cut on slow pion momentum, 
fitting and yield determination in $MM^2$ distribution, and fake leptons. 
The results of these studies are given in Table~I.

(ii) We have studied  the systematic error due to the background subtraction in 
the $\Delta M$ distribution. We included    
true $D^{*+} \to D^0 \pi_s^+$, $D^0\to K^-\pi^+$ decays 
where the  $D^{*+}$ does not come from a signal decay chain in 
the definition of background.
The main source of this background is  $D^{*+} \ell^-$ pairs for which  
the $D^{*+}$ comes from one $\overline B^0$, 
and the lepton is the  primary lepton from another $\overline B^0$. 
This background is suppressed because it occurs only due to  $ B^0-\overline B^0$ mixing.
 A less significant  source  is  $D^{*+} \ell^-$ pairs  for which
the $D^{*+}$  comes from  one $\overline B^0$ or $B^-$ and the  
lepton  is a secondary  lepton from the $\overline{D}$ from the other  $ B^0$ or  $B^+$.
This background is suppressed by the lepton momentum requirement  which  
predominantly selects primary leptons from $B$ decays.
Neither  of these background components   contribute to the peak at $MM^2\approx 0$ 
because the lepton and  slow pion come from different $B$'s. 
We varied the Monte Carlo prediction for  these backgrounds by 20\%  
(based on the conservative estimate of the 
uncertainties in the inclusive $D^{*+}$ and lepton yields, 
the $ B^0-\overline B^0$ mixing parameter, and  the dependence of  $MM^2$ distribution 
shape  on  the $D^{*+}$ momentum  spectrum),  and 
obtained $0.6\%$ variation in final result, which we took as the systematic error.

The rest of the background in the $\Delta M$ distribution is 
combinatoric.  
To estimate the systematic error due to the Monte Carlo simulation of this background 
we substituted the combinatoric part of the Monte Carlo  background shape by an analytic threshold function and obtained the 0.9\% shift  in the final result, which we took as the systematic error.   

 Combining the  errors described  above in (ii) we estimated the systematic error due  to background subtraction in the the  $\Delta M$ distribution to be 1.1\%. 
We have also studied the possible systematic error 
due to the fitting and yield determination in the  $\Delta M$  distribution, and 
the result of this study is given in Table~I. 
 
(iii)  A study has been performed to estimate the systematic error due to the extraction of the 
reconstruction efficiency for $K^-$ and $\pi^+$ tracks from Monte Carlo.
We assigned a 2\% error to the final result (1\% per track).   
As was  mentioned earlier, the  lepton and slow pion reconstruction efficiencies do
not cancel out exactly due to the difference in charged multiplicity between the cases 
$D^0 \to K^-\pi^+$ and $D^0 \to all$. 
To estimate the systematic error due to this effect we extracted the efficiency from Monte Carlo 
forcing  $D^0 \to K^-\pi^+$ when we determine  $N^{incl}_{MC}$.
As a systematic error we took 30\% of the shift in the efficiency obtained using this method and the method actually employed in the analysis.
We have also studied the possible systematic error 
due to the choice of the signal region in the  $\Delta M$  distribution, 
and the result of this study is given in Table~I. 

The systematic errors due to the limited Monte Carlo statistics and the 
continuum subtraction are also given in Table~I.

\begin{table}[htp]
\begin{center}
\begin{tabular}{|c l|c|} 
Quantity  & Possible source     of systematic error     &  Estimate of   Error \\
          &       &  (\%  of final result) \\ \hline \hline 
          
$N^{incl}$&  Background subtraction in $MM^2$  distribution & $2.5\%$  \\ 
          &   Slow pion momentum cut (affects $MM^2$ background shape)    & $1.0\%$  \\
          & Fitting and yield determination                      & $0.6\%$  \\
          &  Fake leptons                                         & $0.2\%$  \\  \hline

$N^{excl}$ &  Background subtraction in $\Delta M$ distribution & $1.1\%$ \\ 
           &  Fitting and yield determination                       & $0.3\%$   \\ \hline
      
$\epsilon$ & $K^- \pi^+$ reconstruction efficiency               & $2.0\%$   \\
          &Choice of signal region in $\Delta M$ distribution & $1.6\%$  \\ 
& Non-exact cancellation of  $\ell$ and  $\pi_s$ reconstruction  efficiencies  &   $1.1\%$        \\ \hline 
&  Monte Carlo  statistics                                        & $1.4\%$  \\ \hline
          &Continuum subtraction                        & $0.1\%$  \\ \hline \hline 

                 & Total             & $4.3\%$   
\end{tabular}
\end{center} 
\label{tab:syst}
\caption{Systematic error summary table.}
\end{table}

In conclusion, we have measured the absolute  branching fraction for 
$D^0\to K^-\pi^+$ decay using a   
$\overline B \to D^{*+}X\ell^-\bar\nu$ tag. We have found 
$Br(D^0\to K^- \pi^+) = [3.81\pm 0.15(stat.)\pm 0.16(syst.)]\%$ \cite{radiation}. 
Our result is consistent with a recent measurement by ALEPH of 
$(3.82 \pm 0.09 \pm 0.11)\%$ \cite{ALEPH-Kpi}, 
\footnote{We took the value before correction for the final state radiation from the $K$ and $\pi$ daughters in the $D^0$ decay.} two measurements by  ARGUS of $(3.41 \pm 0.12 \pm 0.28)\%$ \cite{ARGUS-Kpi} and
of $(4.5 \pm 0.6 \pm 0.4)\%$ \cite{ARGUS-Kpi-partial},
and two measurements by  CLEO of $(3.91 \pm 0.08 \pm 0.17)\%$ \cite{CLEO-Vivek} and of
$(3.69 \pm 0.11 \pm 0.16)\%$ \cite{CLEO-Ed}.
Taking  into account correlations, we combined our result 
with the other two CLEO measurements and found a new
CLEO average value for $Br(D^0\to K^- \pi^+)$ to be
$[3.82\pm 0.07(stat.)\pm 0.12(syst.)]\%$.

We gratefully acknowledge the effort of the CESR staff in providing us with
excellent luminosity and running conditions.
This work was supported by 
the National Science Foundation,
the U.S. Department of Energy,
the Heisenberg Foundation,  
the Alexander von Humboldt Stiftung,
Research Corporation,
the Natural Sciences and Engineering Research Council of Canada, 
the A.P. Sloan Foundation, 
and the Swiss National Science Foundation.

\begin{thebibliography}{99}
\bibitem{charge-conjugate}
Charge conjugate modes are implied throughout this Letter unless otherwise stated.
$\overline B$  can be either $\overline B^0$ or $B^-$.
\bibitem{radiation}
The $K$ and $\pi$ daughters in the $D^0$ decay  can radiate photons in final state. 
This radiation results in a radiative tail of  the $M(K \pi)$ distribution, which we
did not include in the calculation of the efficiency. 
The requirement $|M(K^- \pi^+)-M(D^0)|<35$ MeV$/c^2$ implies 
the effective cut on $\gamma$ energy in the $D^0$ rest frame of $E_{\gamma}^{*}<30$ MeV.
Therefore our  result corresponds to the  value of $Br(D^0 \to K^- \pi^+ (\gamma) )$ with 
the effective cut on $\gamma$ energy in the $D^0$ rest frame of $E_{\gamma}^{*}<30$ MeV.
We compare our measurement with several other measurements which have the $M(K \pi)$ resolution comparable to our resolution,  $\sigma(M(K \pi)) \simeq $ 10 MeV$/c^2$.
\end{thebibliography}
\end{document}